\documentstyle[aps]{revtex}
\input epsf
\def\jepsfbox#1{\typeout{#1} \epsfbox{#1}}

\def\plotone#1{\begin{center} \leavevmode
\epsfxsize=\columnwidth \jepsfbox{#1} \end{center}}

\def\jcite#1#2{#1 \cite{#2}}
\def\ie{{\it i.e.~}}

\def\rmmat#1{{\hbox{\rm #1}}}
\def\rmscr#1{\rmmat{\scriptsize #1}}
\newcommand{\be}{\begin{equation}}
\newcommand{\ee}{\end{equation}}
\newcommand{\ba}{\begin{eqnarray}}
\newcommand{\ea}{\end{eqnarray}}
\def\p{\partial}

\def\pp#1#2{\frac{\p #1}{\p #2}}
\def\figref#1{Fig.~\ref{fig:#1}}
\def\eqref#1{Eq.~\ref{eq:#1}}
\begin{document}
%
\newcommand{\bfi}{{\bf B}} \newcommand{\efi}{{\bf E}}
\newcommand{\kfi}{{\bf k}}
\newcommand{\afi}{{\bf A}}
\newcommand{\lag}{{\Lambda_\rmscr{QED}}}
\newcommand{\dLIII}{{\frac{\partial^3\lag}{\partial I^3}}}
\newcommand{\dLII}{{\frac{\partial^2\lag}{\partial I^2}}}
\newcommand{\dLI}{{\frac{\partial \lag}{\partial I}}}
\newcommand{\dLKKK}{{\frac{\partial^3 \lag}{\partial K^3}}}
\newcommand{\dLKK}{{\frac{\partial^2 \lag}{\partial K^2}}}
\newcommand{\dLK}{{\frac{\partial \lag}{\partial K}}}
\newcommand{\dLIK}{{\frac{\partial^2 \lag}{\partial I \partial K}}}
\newcommand{\dLJJ}{{\frac{\partial^2 \lag}{\partial J^2}}}
\newcommand{\dLIJ}{{\frac{\partial^2 \lag}{\partial I \partial J}}}
\newcommand{\dLJ}{{\frac{\partial \lag}{\partial J}}}
\newcommand{\dLJJJ}{{\frac{\partial^3 \lag}{\partial J^3}}}
\newcommand{\dLIJJ}{{\frac{\partial^3 \lag}{\partial I \partial J^2}}}
\newcommand{\dLIIJ}{{\frac{\partial^3 \lag}{\partial I^2 \partial J}}}
\title{Nonlinear QED Effects in Strong-Field Magnetohydrodynamics}
\author{Jeremy S. Heyl}
\address{Theoretical
Astrophysics 130-33, California Institute of Technology,
Pasadena, California 91125}
\author{Lars Hernquist}
\address{Lick Observatory,
University of California, Santa Cruz, California 95064, USA}
\maketitle
\begin{abstract}
We examine wave propagation and the formation of shocks in strongly
magnetized plasmas by applying a variational technique and the method
of characteristics to the coupled magnetohydrodynamic (MHD) and
quantum-electrodynamic (QED) equations of motion.  In sufficiently
strong magnetic fields such as those found near neutron stars, not
only is the plasma extremely relativistic but the effects of QED must
be included to understand processes in the magnetosphere.  As
\jcite{Thompson \& Blaes}{Thom98} find, the fundamental modes in the
extreme relativistic limit of MHD coupled with QED are two oppositely
directed Alfv\'{e}n modes and the fast mode.  QED introduces nonlinear
couplings which affect the propagation of the fast mode such that
waves traveling in the fast mode evolve as vacuum electromagnetic
ones do in the presence of an external magnetic field
\cite{Heyl98shocks}.  The propagation of a single Alfv\'{e}n mode is
unaffected but QED does alter the coupling between the Alfv\'{e}n
modes.

This processes may have important consequences for the study of
neutron-star magnetospheres especially if the typical magnetic field
strength exceeds the QED critical value ($B_\rmscr{QED}
\approx 4.4 \times 10^{13}$~G) as is
suspected for soft-gamma repeaters and anomalous X-ray pulsars.
\end{abstract}

\section{Introduction}

Ultrarelativistic plasmas play an important role in energy
transmission in many astrophysical settings including neutron-star
magnetospheres, black hole accretion disks and the sources of
gamma-ray bursts.  Additionally, in the case of neutron stars, the
magnetic field may exceed the QED critical value ($\approx 4.4 \times
10^{13}$~G) and vacuum corrections may affect dynamics of the
electromagnetic field.

In relativistic field theory, it is natural to study dynamics using a
Lagrangian formulation.  In the case of QED, the one-loop vacuum
corrections may be summarized by an effective Lagrangian which
includes both the classical Lagrangian and the consequences of virtual
pairs \cite{Heis36,Weis36,Schw51,Heyl97hesplit}.  If the separation of
the charges comprising the plasma can be neglected, the dynamics of
the plasma may be treated using magnetohydrodynamics.  Again the
determination of a suitable Lagrangian expedites the relativistic
treatment of magnetohydrodynamics \cite{Acht83,Thom98}.

\section{The Action}

The Lagrangian derived by \jcite{Achterberg}{Acht83} may be simplified
dramatically if the inertia of the charge carriers can be neglected,
\ie in the extreme relativistic limit.  \jcite{Thompson \& Blaes}{Thom98}
present two formulations for magnetohydrodynamics (MHD) appropriate in
this limit.  Furthermore, they also examine the limit where $\omega^2,
k_\perp^2 \ll eB$.  In this limit, the passing MHD wave cannot excite
the charge carriers into the second Landau level, so they are
effectively trapped along a single field line.  The fermion fields are
restricted to $1+1$ dimensions, and they may be treated using the
technique of bosonization.  The fermion fields are replaced by a
four-dimensional axion field ($\theta$) which enforces the MHD
condition, \ie $\efi \cdot \bfi = 0$.

\jcite{Thompson \& Blaes}{Thom98} obtain the simple action for
the electromagnetic field in the presence of the relativistic plasma:
\be
S'' = \int d^4 x \left [ -\frac{1}{4} I
	+ \frac{e^2}{2} \theta J \right ].
\ee
We have the following additional definitions,
\be
I = F_{\mu\nu} F^{\mu\nu} \rmmat{~and~}
J = {\cal F}^{\mu\nu} F_{\mu\nu}
\ee
where ${\cal F}^{\mu\nu}$ is the dual to the field tensor given by
\be
{\cal F}^{\mu\nu} = \frac{1}{2} \epsilon^{\rho\lambda\mu\nu} F_{\rho\lambda}.
\ee
$\epsilon^{\rho\lambda\mu\nu}$ is the completely antisymmetric
Levi-Civita tensor.

Here we will examine the modified action where
\be
S'' = \int d^4 x \left [ \lag(I,K) + \frac{e^2}{2} \theta J \right ]
\ee
where $K=-J^2$.  To maintain the $CP$ and Lorentz invariance of QED,
its effective Lagrangian $\lag$ must be a function of the field
scalars
\ba
I &=& 2 \left ( |\bfi|^2 - |\efi|^2 \right ), \\
K &=& - \left ( 4 \efi \cdot \bfi \right )^2,
\ea
rather than of the pseudoscalar $J=-4 \efi \cdot \bfi$.

\section{Wave propagation}

\begin{figure}
\plotone{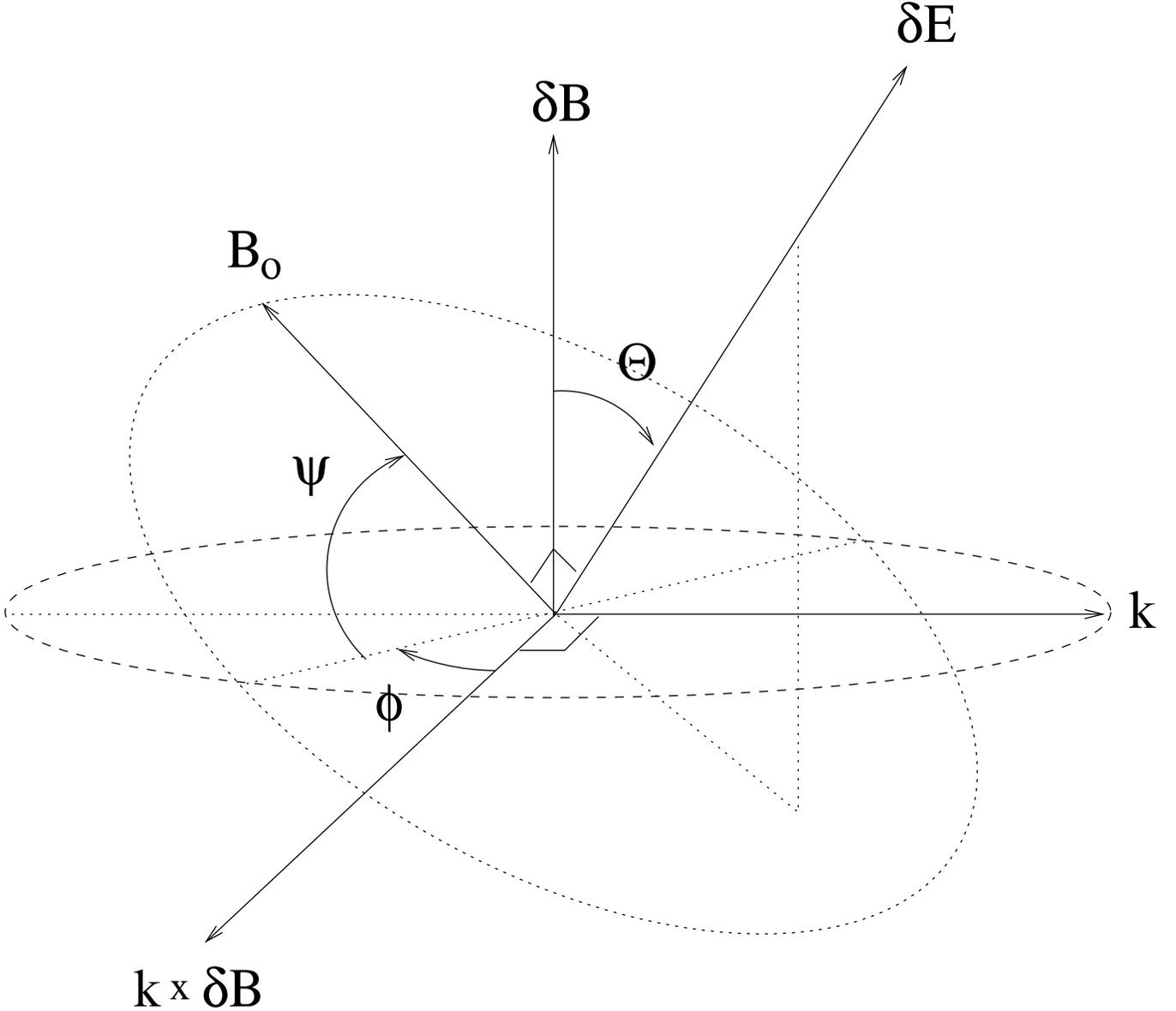}
\caption{The configuration of the MHD wave.  N.B.: in the configuration
depicted, all three Euler angles are less than zero.}
\label{fig:mhdgeom}
\end{figure}

To study the propagation of waves through the plasma we will use the 
formalism of \jcite{Heyl \& Hernquist}{Heyl98shocks}.  We use the
results of \jcite{Thompson \& Blaes}{Thom98} to describe the
traveling modes.  Specifically we designate the external magnetic
field by $\bfi_0$ and the electric and magnetic fields associated with
the wave by $\delta \efi$ and $\delta \bfi$ respectively.  We also
have the constraints $\delta \efi \cdot \bfi_0 = 0$ and $\kfi \cdot
\delta \bfi = 0$ where $\kfi$ is the wave vector.  \figref{mhdgeom}
depicts the geometry of the propagating wave and defines the three
Euler angles $\Psi, \Theta$ and $\phi$ used to describe its
configuration \cite{Gold80}.  We will take the $x$, $y$ and $z-$axes
to be aligned with $\delta \bfi \times \kfi$, $\kfi$ and $\delta \bfi$
respectively.

The definitions allow us to calculate the invariants 
\ba
I &=& 2 \left [ B_0^2 + (\delta B)^2 + 
	2 B_0 \delta B \sin \Theta \sin \Psi - ( \delta E)^2 \right ], \\
J &=& -4  (\delta B) (\delta E) \cos \Theta.
\ea

\subsection{The Lagrange Condition}

To calculate the equations of motion of the wave we assume that the
wave fields $\delta \bfi$ and $\delta \efi$ and the axion field
$\theta$ are dynamic while the external magnetic field $B_0$ is
static.  Varying the action with respect to the axion field yields,
\be
\pp{L}{\theta} = 0 \Rightarrow J = -4 \efi \cdot \bfi = 0.
\ee
The field $\theta$ acts as a Lagrange multiplier to enforce the MHD
condition.  The equations of motion for the fields $\delta \efi$ and
$\delta \bfi$ are more complicated than in the vacuum case
\cite{Heyl98shocks} because here the relationship between the wave fields and
their potential is more complicated.

\subsection{The four-potential}

We will designate the potential of the wave by the four-vector 
$\delta A_\mu = [ \delta A_t, \delta \afi(x,y,z) ]$.  To within a gauge
transformation the vector potential is given by $(\delta \afi)_x=-\psi(y,t)$
and $(\delta \afi)_y=(\delta \afi)_z=0$.

Let us now examine the electric field,
\be
\delta \efi = - {\bf \nabla} (\delta A_t) - \pp{\delta \afi}{t}.
\ee
A portion of the electric field may be due to the scalar potential.
We define
\be
\delta \efi_\rmscr{irr} = \delta \efi + \pp{\delta \afi}{t}.
\ee
Let us insist that the direction of $\delta \efi$ is constant in time
and its magnitude is a function of $y$ and $t$ only.  Therefore, we
have $\delta \efi= \delta E(y,t) {\bf x}$ and calculate the curl of the
irrotational component to obtain that $(\delta \efi)_z=0$ (\ie $J=0$,
$\cos \Theta=0$), unless we have $\delta \efi$ constant
throughout space.   We also find that
\be
(\delta \efi)_x = \psi_{,t} + f(x,z,t).
\ee
However, since $(\delta \efi)_x$ and $\psi$ are functions of $y$ and $t$ we
find that $f(x,z,t)$ depends only on $t$.  $(\delta \efi)_y$ is
still unconstrained.  However we do know that $(\delta \efi)_y = R
(\delta \efi)_x$.  We can immediately deduce the scalar potential obtaining
\be
\delta A_t =  - y R  f(t) - R \int_0^y \psi_{,t} (u,t) d u.
\ee
The first group of terms  yields a constant background electric field
in space which can vary in time.  We drop this extra field and obtain
the following four-potential,
\ba
\delta A_t &=& R \int_0^y \psi_{,t} (u,t) d u \\
\delta A_y &=& \delta A_z = 0 \\
\delta A_x &=& -\psi(y,t)
\ea
where
\be
R = \frac{(\delta \efi)_y}{(\delta \efi)_x} = -\cot \phi.
\ee
The four-potential derived here is more general than that calculated
by \jcite{Heyl \& Hernquist}{Heyl98shocks} in that it contains an
electric field such that $\kfi \cdot \delta \efi \neq 0$ which
complicates the derivation of the equations of motion.  In summary we
have
\be
(\delta \bfi)_x = (\delta \bfi)_y = (\delta \efi)_z = 0
\ee
and
\be
(\delta \bfi)_z = \psi_{,y}, (\delta \efi)_x = \psi_{,t},
\rmmat{~and~} (\delta \efi)_y = R \psi_{,t}.
\ee

\subsection{Equations of Motion}

The equations of motion are obtained by attempting to minimize the
action with respect to the electromagnetic potential and the axion
field.  We found earlier that the axion field acts as a Lagrange
multiplier to enforce the MHD condition.  For the potential we obtain
the following expression
\be
\partial_\mu \pp{L}{(\delta A_{\nu,\mu})} = 0.
\ee
Substituting the MHD action yields,
\be
\partial_\mu \left [ \left ( 2 e^2 \theta - 8 J  \dLK \right )
{\cal F}^{\mu\nu} + 4 \dLI F^{\mu\nu} \right ] = 0.
\label{eq:Lagrange}
\ee
We see that the axion field and QED alter the equations of motion for the
electromagnetic field,
\be
\partial_\mu F^{\mu\nu} = J^\nu.
\ee
where
\be
J^\nu = -\frac{1}{\cal B} \left ( {\cal F}^{\mu\nu} \partial_\mu {\cal
A} + F^{\mu\nu} \partial_\mu {\cal B} \right )
\ee
with
\ba
{\cal A} &=& 2 e^2 \theta - 8 J  \dLK \\
{\cal B}  &=& 4 \dLI.
\ea
If we neglect the effects of QED on the dynamics we obtain that
\ba
\rho &=& 2 e^2 \bfi \cdot {\bf \nabla} \theta 
\label{eq:mhdrho} \\
{\bf J} &=& 2 e^2 \left ( \efi \times {\bf \nabla} \theta - \theta_{,t} \bfi \right ) .
\label{eq:mhdJ}
\ea
If we specialize to the geometry and potential described in the
previous subsections and eliminate the axionic degrees of freedom from
the equations, we find that remaining components of
\eqref{Lagrange} can be written as
\ba
a_1 \psi_{,yy} + b_1 \psi_{,yt} + c_1 \psi_{,tt} &=& 0 
\label{eq:quad1} \\
a_3 \psi_{,yy} + b_3 \psi_{,yt} + c_3 \psi_{,tt} &=& 0
\label{eq:quad3} 
\ea
where
\ba
a_1 &=& -\left [ 1 + Q \left (\psi_{,y} +
B_{0z} \right )^2 \right ] 
\label{eq:a1def}\\
b_1 &=& - Q \psi_{,t} \left ( \psi_{,y} + B_{0z} \right
) \left ( \frac{R B_{0x}}{B_{0y}} - 2 - R^2 \right ) 
\label{eq:b1def}
\\
c_1 &=& Q  \psi_{,t}^2 \left (1 + R^2 \right )
\left ( \frac{R B_{0x}}{B_{0y}} - 1 \right ) - \frac{R B_{0x}}{B_{0y}}
+ 1
\label{eq:c1def}\\
a_3 &=& Q \left ( \psi_{,y} + B_{0z} \right ) \left (
\psi_{,t}^2 \frac{R}{B_{0y}} + B_{0x} \right ) 
\label{eq:a3def}\\
b_3 &=&    \frac{\psi_{,t} R}{B_{0y}} \left \{ 1 
-Q
\left [   \left (1+R^2 \right )
\left ( \psi_{,t}^2  + B_{0x} \frac{B_{0y}}{R} \right ) +
 \left ( \psi_{,y} + B_{0z} \right )^2 \right ]
\right \} 
\label{eq:b3def}
\\
c_3 &=& \frac{(\psi_{,y}+B_{0z}) R}{B_{0y}} \left [
Q \psi_{,t}^2 \left ( 1 + R^2 \right ) - 1
\right ]
\label{eq:c3def}
\ea
where
\ba
B_{0x} &=& B_0 \cos \Psi \cos \phi \\
B_{0y} &=& B_0 \cos \Psi \sin \phi \\
B_{0z} &=& B_0 \sin \Theta \sin \Psi \\
Q &=& 16 \dLII \left / \dLI \right . .
\ea

\subsection{The Classical Limit}

If we neglect the effects of QED in the equations above, we can derive
the wave equations in relativistic MHD.  Specifically, we take
$Q = 0$ (in the classical limit $\partial \lag / \partial I=-1/4$) to
obtain
\ba
v^{-2} \psi_{,tt} - \psi_{,yy} &=& 0
\label{eq:mhd1} \\
\frac{R}{B_{0y}} \left ( \psi_{,t} \psi_{,yt} - \psi_{,y} \psi_{,tt} -
B_{0z} \psi_{,tt} \right ) &=& 0
\label{eq:mhd3}
\ea
where
\be
v^{-2} = 1 - \frac{R B_{0x}}{B_{0y}}.
\ee
\eqref{mhd1} is satisfied by
\be
\psi(y,t) = \psi(y \pm v t).
\ee
This equation also satisfies \eqref{mhd3} if $R=0$ or
$B_{0z}=0$; consequently, MHD supports fully nonlinear modes if
${\delta \efi} \cdot \kfi=0$ (fast modes)
or ${\delta \bfi} \cdot \bfi_0=0$ (Alfv\'{e}n modes).

For the fast modes, we have $R=0$ and a dispersion relation of 
$\omega^2=k^2$.  For the Alfv\'{e}n modes, we find that $\bfi_0, \delta \efi$
and $\kfi$ all lie in the plane perpendicular to $\delta \bfi$.  Since
$\delta \efi \cdot \bfi_0=0$ we have
\ba
R = \frac{(\delta \efi)_y}{(\delta \efi)_x} = -\frac{B_{0x}}{B_{0y}} 
\Rightarrow v_\rmscr{A}^2 = \frac{1}{1 + R^2} = 
\sin^2 \phi,
\ea
yielding a dispersion relation of $\omega = \pm k_{B_0}$.  $k_{B_0}$
is the component of $\kfi$ directed along the external magnetic field.

However, even if ${\delta \bfi} \cdot \bfi_0=0$ unless $\bfi_0$ and
$\kfi$ are parallel (\ie $R=0$),
\be
\psi(y,t) = \psi_1(y + v_\rmscr{A} t) + \psi_2(y - v_\rmscr{A} t)
\label{eq:twoawaves}
\ee
does not satisfy \eqref{mhd3}.  Two oppositely traveling Alfv\'{e}n modes
will interact through the cross term,
\be
\frac{R v_\rmscr{A}^2}{B_{0y}} \left ( \psi_{1,yy} \psi_{2,y} + \psi_{2,yy} \psi_{1,y} \right ) = 0.
\ee
It is also straightforward to derive equations \eqref{mhd1} and \eqref{mhd3} 
from \eqref{mhdrho} and \eqref{mhdJ} using Maxwell's equations.

\subsection{The Effects of QED}

QED introduces several additional terms into the equations of motion.
By restricting the geometry of the wave, we can still satisfy
\eqref{quad3} identically.  

\subsubsection{Fast Modes}
\label{sec:fast}

First, we examine the fast modes 
which have both $R=0$ and $B_{0x}=0$ and 
therefore satisfy \eqref{quad3} even when QED effects are
included.  For these modes, we obtain
\ba
b_1 &=& 2 Q \psi_{,t} \left ( \psi_{,y} + B_{0z}
\right ) \\
c_1 &=& 1 - Q \psi_{,t}^2.
\ea
Because $a_1$ does not depend on $R$ or $B_{0x}$ it is still given by 
\eqref{a1def}.  As in \jcite{Heyl \& Hernquist}{Heyl98shocks} 
we expand the coefficients to first order in the fields,
\ba
a_1 &=& a_{1,0} + a_{1,B} \psi_{,y} + a_{1,E} \psi_{,t} + {\cal O} (\delta B^2) \\
b_1 &=& b_{1,0} + b_{1,B} \psi_{,y} + b_{1,E} \psi_{,t} + {\cal O} (\delta B^2) \\
c_1 &=& c_{1,0} + c_{1,B} \psi_{,y} + c_{1,E} \psi_{,t} + {\cal O} (\delta B^2).
\ea
Notice the sign change relative to \jcite{Heyl \&
Hernquist}{Heyl98shocks}.
In the previous work, we selected a gauge where $(\delta E)=-\psi_{,t}$.
We obtain
\ba
a_{1,0} &=& -4 \left [ 4 B_{0z}^2 \dLII + \dLI \right ] \\
a_{1,B} &=& -16 \left [ 3 \dLII B_{0z} + 4 \dLIII B_{0z}^3 \right ] \\
b_{1,E} &=& 32 \dLII B_{0z} \\
c_{1,0} &=& 4 \dLI \\
c_{1,B} &=& 16 \dLII B_{0z},
\ea
and 
\be
a_{1,E}=b_{1,0}=b_{1,B}=c_{1,E}=0.
\ee
If we substitute these expansions back into \eqref{quad1} we obtain an
identical result to that of \jcite{Heyl \& Hernquist}{Heyl98shocks} in
the limit of $K=0$.  This is not particularly surprising because the
fast modes do not carry current and cannot excite the axion field.
Furthermore, the results of \jcite{Heyl \& Hernquist}{Heyl98shocks}
may be applied directly to understand the evolution of the fast modes
including the effects of QED to one-loop order.

\jcite{Heyl \& Hernquist}{Heyl98shocks} found that a electromagnetic
wave traveling through a strongly magnetized vacuum will develop
discontinuities after traveling a distance proportional to its
wavelength and amplitude.  Furthermore, the opacity to shocking peaks
near the critical field ($\approx 4.4 \times 10^{13}$~G).  After the
discontinuity forms, the energy of the wave is quickly dissipated most
likely as electron-positron pairs.  The results of this section show
that the presence of the plasma does not affect the development of
this nonlinearity for the fast modes; therefore, even in the
plasma-filled magnetosphere surrounding a neutron star, one would
expect shocks to develop as waves in the fast mode propagate.

\subsubsection{Alfv\'{e}n Modes}
\label{sec:Alfven}

The treatment of the current-carrying modes within QED is more
subtle.  We can use the results that $B_{0z}=0$ and $R=-B_{0x}/B_{0y}$
to simplify the equations,
\ba
a_1 &=& -1 - Q \psi_{,y}^2 
\label{eq:a1shear} \\
b_1 &=& 2 Q \psi_{,t} \psi_{,y} 
\left ( 1 + R^2 \right ) 
\label{eq:b1shear} \\
c_1 &=& - Q \psi_{,t}^2 \left (1 + R^2 \right )^2
+ 1 + R^2
\label{eq:c1shear} 
\\
a_3 &=& \frac{\psi_{,y} R}{B_{0y}} Q  \left (
\psi_{,t}^2 - B_{0y}^2 \right ) 
\label{eq:a3shear} \\
b_3 &=&  \frac{\psi_{,t} R}{B_{0y}} \left \{ 1 
-Q
\left [   \left (1+R^2 \right )
\left ( \psi_{,t}^2  - B_{0y}^2 \right ) +
 \psi_{,y}^2 \right ]
\right \} 
\label{eq:b3shear} 
\\
c_3 &=& \frac{\psi_{,y} R}{B_{0y}} \left [
 Q \psi_{,t}^2  \left ( 1 + R^2 \right ) - 1
\right ].
\label{eq:c3shear} 
\ea

To proceed we combine \eqref{quad1} and \eqref{quad3} into a single
relation by eliminating $\psi_{,yt}$,
\be
-v^2 \psi_{,yy} + \psi_{,tt} = 0
\label{eq:qeda}
\ee
where
\ba
v^2 &=& -\frac{a_1 b_3 -  a_3 b_1}{b_3 c_1 - b_1 c_3}
\\
&=& \frac{1}{1+R^2} \frac{1 - Q \psi_{,y}^2}{1-Q\psi_{,t}^2
\left(1+R^2\right)}.
\ea
This elimination is impossible in pure MHD (\ie $Q=0$) since the term
$\psi_{,yt}$ appears only in the second of the equations of motion 
(\eqref{mhd3}).

Let us attempt to use a simple wave solution similar to that of MHD,
\be
\psi = \psi \left ( y \pm v_\rmscr{A,QED} t \right ).
\ee
If we substitute this into \eqref{qeda}, we find
\be
v_\rmscr{A,QED}^2 = \frac{1}{1+R^2} 
\frac{1 - Q \psi_{,y}^2}{1-Q v_\rmscr{A,QED}^2 \psi_{,y}^2 \left(1+R^2\right)}
\ee
which is satisfied by
\be
v_\rmscr{A,QED}^2 = v_\rmscr{A}^2 = \frac{1}{1+R^2}.
\ee
We find that a single Alfv\'{e}n mode does not suffer any nonlinearities due to 
QED if the QED effective Lagrangian is invariant with respect to gauge 
and Lorentz transformations and
\be
\dLI \neq 0.
\ee
\eqref{qeda} like \eqref{mhd3} indicates that two oppositely
traveling Alfv\'{e}n modes will interact.  If we substitute \eqref{twoawaves} 
into the equations of motion for Alfv\'{e}n waves including QED we obtain
\ba
Q \psi_{2,y}^2 \psi_{1,yy} + \left \{ 1 \Longleftrightarrow 2 \right \}  &=& 0 \\
\frac{R v_\rmscr{A}^2}{B_{0y}} \psi_{2,y} \psi_{1,yy} \left \{ 1 + 
Q \left [  B_{0}^2 + 2 \psi_{2,y} \left ( \psi_{1,y}  - \psi_{2,y} \right ) 
  \right ] \right  \}
+ \left \{ 1 \Longleftrightarrow 2 \right \}  &=& 0
\ea
where $\left \{ 1 \Longleftrightarrow 2 \right \}$ designates the same
terms repeated with the functions $\psi_1$ and $\psi_2$ swapped to
obtain a symmetric sum. 

To lowest order in the wave fields we find that QED does not affect the
coupling between the Alfv\'{e}n waves.  However, at higher order, vacuum
processes introduce several new interaction terms.

\section{Conclusions}

We find that QED affects the propagation of magnetohydrodynamic fast
modes traveling through an external magnetic field.  The induced
nonlinearity is identical to that suffered by electromagnetic
radiation traveling through an external magnetic in the absence of a
plasma.  For the Alfv\'{e}n modes, we find that QED introduces
additional couplings between oppositely directed waves, but that
because of the gauge and Lorentz invariance of QED, no nonlinearities
manifest themselves in the propagation of a single Alfv\'{e}n mode.

These new nonlinear processes emerge in regions where the magnetic
field strength is comparable to or exceeds $B_\rmscr{QED} \approx 4.4
\times 10^{13}$~G.  Both anomalous X-ray pulsars (AXPs) and soft-gamma
repeaters (SGRs) circumstantially exhibit such strong magnetic fields
\cite{Vasi97b,Heyl97kes,Heyl98decay,Thom96,Kouv98}; consequently,
these nonlinear processes specific to intense magnetics fields may
play a important role in the magnetospheres surrounding these objects.

\acknowledgements
We would like to thank Peter Goldreich and Roger Blandford for useful
discussions.  J.S.H. would also like to acknowledge a Lee A. DuBridge
Postdoctoral Scholarship and Cal Space Grant CS-12-97.


\end{document}